\documentclass[11pt,fleqn]{article}

\usepackage{amsmath} 
\usepackage{amsfonts}
\usepackage{graphicx}
\usepackage{dcolumn}
\usepackage{upgreek}
\usepackage{latexsym,slashed}
\usepackage{amssymb,amsfonts}
\usepackage{cancel}
\usepackage{bm} 
\usepackage{amsmath,mathrsfs} 

\newcommand{\pacs}[1]{\smallskip\noindent{\sl PACS numbers: \hspace{0.3cm}#1}\par\bigskip\rm}

\DeclareMathOperator{\sech}{sech}

\newcommand{\address}[1]{\begin{center}\large #1\end{center}}

\def\beq{\begin{eqnarray}}
\def\eeq{\end{eqnarray}}
\def\T{\tau}
\def\p{\partial}

\def\R{{\hbox{{\rm I}\kern-.2em\hbox{\rm R}}}} 
\def\H{{\hbox{{\rm I}\kern-.2em\hbox{\rm H}}}} 
\def\N{{\hbox{{\rm I}\kern-.2em\hbox{\rm N}}}} 
\def\C{{\ \hbox{{\rm I}\kern-.6em\hbox{\bf C}}}} 
\def\Z{{\hbox{{\rm Z}\kern-.4em\hbox{\rm Z}}}} 
 
\newcommand{\ack}[1]{\par\section*{Acknowledgement} #1}

\catcode`\@=11
\@addtoreset{equation}{section}

\begin{document}
\tolerance=5000

\title{Unruh-DeWitt detector and the interpretation of the horizon temperature in spherically symmetric dynamical space-times} 

\author{Giovanni Acquaviva$\,^{(a)}$\footnote{acquaviva@science.unitn.it},
Roberto~Di~Criscienzo$\,^{(a)}$\footnote{rdicris@science.unitn.it},
Luciano~Vanzo$\,^{(a)}$\footnote{vanzo@science.unitn.it} and
Sergio~Zerbini$\,^{(a)}$\footnote{zerbini@science.unitn.it}}
\date{}
\maketitle
\address{$^{(a)}$ Dipartimento di Fisica, Universit\`a di Trento \\
and Istituto Nazionale di Fisica Nucleare - Gruppo Collegato di Trento\\
Via Sommarive 14, 38123 Povo, Italia}
\medskip \medskip

\begin{abstract}
In the paper, the temperature associated with a dynamical spherically symmetric black hole or with a cosmological horizon is investigated from the point of view of a point-like detector. First, 
we briefly review the Hamilton-Jacobi tunneling method for a generic dynamical spherically symmetric space-time, and present two applications of the tunneling method. Then, we apply a well-known relativistic quantum theoretical 
technique, namely the Unruh-DeWitt detector formalism for a conformally coupled scalar field in a generic FRW space-time. As an application, for the generic static black hole case and the FRW de Sitter case, making use of peculiar
Kodama observer trajectories, the tunneling semiclassical results are fully recovered, automatically corrected by Tolman factors. Some remarks on the temperature of FRW universe are presented. For more general spaces interpolating de Sitter space with the Einstein-de Sitter universe a second set of poles is present, whose exact role remains to be clarified, plus an extra fluctuating term describing the way equilibrium is reached, similarly to de Sitter space. The simple thermal interpretation found for de Sitter space is lost and forces, at a same time, a different quantum interpretation of the horizon surface gravity for the cosmological FRW models.
\end{abstract}


\section{Introduction}

It is well known that Hawking radiation \cite{Haw} is considered one of the most important prediction of Quantum Field Theory in 
curved space-time. Several derivations of this effect have been proposed \cite{dewitt,BD,wald,fulling,igor} and recently the search for ``experimental'' verification making use of analogue models has been pursued by many investigators (see for example \cite{Unruh:2008zz,Barcelo:2005fc}). 

In 2000 Parikh and Wilczek \cite{PW}(see also \cite{visser}) introduced the so-called tunneling approach
for investigating Hawking radiation. Here, we shall firstly review a variant of their method, called Hamilton-Jacobi tunneling method \cite{Angh,tanu,mann} and also \cite{obr}. This method is covariant and can be extended to the dynamical case 
\cite{bob07,bob08,sean09,bob09}, and to the study of decay of massive particles and particle creation by naked singularities \cite{bob10}. 
In their approach, Parikh \& Wilczek made a clever use of the Painlev\'e stationary gauge for 4-dimensional Schwarzschild black hole 
\beq
ds^2=-\left(1-\frac{2mG}{r}\right)dT^2+ 2\sqrt{\frac{2Gm}{r}}drdT+dr^2+r^2d\Omega^2 \,,
\eeq
which is regular on the trapping horizon $r_H=2mG\;$\footnote{The global event horizon is $r_0<r_H$ for an evaporating black hole and $r_{0}>r_{H}$ for an accreting black hole, as can be seen from the equation of radial null rays, $\dot{r}_{0}=1-\sqrt{r_{H}/r_{0}}$.}. This is one of the key points since the use of singular gauges, as the Schwarzschild gauge, leads, in general, to ambiguities and it is useless in the dynamical case which we are interested in.\\
The second merit we can address to the Parikh \& Wilczek work was a treatment of the back-reaction on the metric, based on energy conservation. In the following, we shall limit to leading term results and neglect the back-reaction. However, it may be worth to recall that in the limit where the number of emitted quanta is reasonably large back-reaction effects can be accounted for by assuming that the mass parameter $m$ is a continuous function of time $T$. Penrose's diagrams for this more general case have been determined too, e.g. in \cite{Lindesay:2010uv,Brown:2009by}.

The Hamilton-Jacobi method is reasonably simple, even though subtleties are present (see, for example \cite{seanrep}). It is based on the computation of the classical action $I$ along a trajectory starting slightly behind the trapping horizon but ending in the bulk, and the associated WKB approximation ($c=1$)
\beq
\mbox{Amplitude} \propto e^{i \frac{I}{\hbar}}\, .
\eeq 
For an evaporating black hole such a trajectory would be classically forbidden since at the trapping horizon photons are only momentarily at rest, whence $dr/dT<0$ inside the horizon\footnote{It means the photon must go back in time $T$ to escape the horizon.}. The related semiclassical emission rate reads 
\beq
\Gamma \propto |\mbox{Amplitude}|^2 \propto 
e^{-2\frac{\Im I}{\hbar}} \,. 
\eeq
with $\Im$ standing for the imaginary part. In the tunneling across the horizon, the 
imaginary part of the classical action $I$ stems from the interpretation of a formal horizon divergence, and in 
evaluating it, one has to make use of Feynman's prescription related to a simple pole in integration such as
\beq
\int_{0_-}^a dx \frac{f(x)}{x}\, \rightarrow\int_{0_-}^a dx \frac{f(x)}{x-i0}=i\pi f(0)+\mbox{real stuff}\,.
\eeq
This corresponds to the choice of suitable boundary conditions in quantum field theory approach.

We may anticipate that in the WKB approximation of the tunneling probability, one asymptotically gets a Boltzmann factor, in which an energy $\omega$ appears, i.e.
\beq
\Gamma \propto e^{- \frac{\beta}{\hbar} \omega}\,.
\eeq
It is crucial in our approach that the argument of the exponent be a coordinate scalar (invariant quantity), since otherwise no physical meaning can be addressed to $\Gamma$. In particular, in the Boltzmann factor, $\beta$ and the energy $\omega$ have to be separately coordinate scalars, otherwise again no invariant meaning could be given to the quantity $\frac{\beta}{\hbar}$. In the static case, we interpret $T=\frac{\hbar}{\beta}$ as the horizon temperature. \\
%
In the cosmological case on the other hand, one can still have a trapping horizon despite the absence of collapsed matter simply as a result of the expansion of the universe. As for an excreting black hole, this too is represented by a time-like hypersurface. Similarly, an approximate notion of temperature can be associated to such horizons based on the existence of a surface gravity and again the tunneling method gives a non vanishing amplitude having the Boltzmann form. However we will see that a comoving monopole detector seems to react to the expansion in a different, non ``Boltzmannian'' way, while reaching thermal equilibrium (or better, detailed balance conditions) only asymptotically for large times. 

The paper is organized in the following way: in Section 2, the Kodama-Hayward formalism and the Hamilton-Jacobi method are summarized; then, in Section 3, we consider the tunneling method for a generic static black hole in Kruskal-like gauge. In Section 4, the same formalism is applied to Friedmann-Robertson-Walker (FRW) spatially flat space-times. A formula for the de Sitter temperature which is valid in any coordinate patch is provided. Section 5, contains a discussion on the Unruh-DeWitt detector which is introduced in order to confirm through quantum field theoretic techniques the results of previous sections. In particular, a quite general formula for the response function per unit proper time is obtained and two relevant applications are presented. Section 6 contains a further example, consisting of a $(\Omega_{m},\Omega_{\Lambda})$ cosmological model, where only approximate results can be obtained. Some toy models are presented and discussed. Concluding remarks follow at the end of the paper.

We use the metric signature $(-,+,+,+)$; Greek indices run over $0$ to $3$ while mid-Latin as $i,j$ only over $0$ and $1$. We use Planck units in which $c=\hbar=G=k_B=1$.

\section{The Kodama-Hayward formalism and Hamilton-Jacobi tunneling method}

In order to treat the spherically symmetric dynamical case, the use of invariant
quantities plays a crucial role \cite{sean09,bob09,kodama,sean}. Here we review the general formalism.

To begin with, let us recall that any spherically 
symmetric metric can locally be expressed in the form
\beq
\label{metric}
ds^2 =\gamma_{ij}(x^i)dx^idx^j+ R^2(x^i) d\Omega^2\,,\qquad i,j \in \{0,1\}\;,
\eeq
where the two-dimensional metric
\beq d\gamma^2=\gamma_{ij}(x^i)dx^idx^j
\label{nm}
\eeq
is referred to
as the normal metric, $\{x^i\}$ are associated coordinates and
$R(x^i)$ is the 
areal radius, considered as a scalar field in the two-dimensional
normal space. 
A relevant scalar quantity in the reduced normal space is 
\beq
\chi(x)=\gamma^{ij}(x)\partial_i R(x)\partial_j R(x)\,, \label{sh} 
\eeq 
since the dynamical trapping horizon, if it exists, is located in
correspondence of 
\beq 
\chi(x)\Big\vert_H = 0\,, \label{ho} 
\eeq
provided that $\partial_i\chi\vert_H \neq 0$.
The Misner-Sharp gravitational energy is defined by
\beq
E_{MS}(x)=\frac{1}{2} R(x)\left[1-\chi(x) \right]\,. \label{MS}
\eeq
This is an invariant quantity on the normal space. Note also that, on
the horizon, $E_{MS}\vert_H =\frac{1}{2} R_H \equiv M$. Furthermore, one can
introduce the Hayward surface gravity associated with this dynamical
horizon, given by the normal-space scalar 
\beq
\kappa_H=\frac{1}{2}\Box_{\gamma} R \Big\vert_H\,. \label{H} 
\eeq 
Recall that, in the spherical symmetric dynamical case, it is possible
to introduce 
the Kodama vector field $\mathcal K$, with $(\mathcal K^\alpha G_{\alpha\beta})^{;\beta}
=0$ that can be taken as its defining property. 
Given the metric (\ref{metric}), the Kodama vector components are
\beq 
\mathcal K^i(x)=\frac{1}{ \sqrt{-\gamma}}\,\varepsilon^{ij}\partial_j R\,,
\qquad \mathcal K^\theta=0=\mathcal K^\varphi \label{ko} \;. 
\eeq 
We may introduce the Kodama trajectories, and related Kodama observer, by means of integral lines of Kodama vector
\beq
\frac{d\, x^i}{d \lambda}=\mathcal K^i= \frac{1}{ \sqrt{-\gamma}}\,\varepsilon^{ij}\partial_j R\,. 
\label{ko1}
\eeq
As a result,
\beq
\frac{d\, R(x(\lambda))}{d \lambda}= \partial_i R\frac{d\, x^i}{d \lambda}=
\frac{1}{ \sqrt{-\gamma}}\,\varepsilon^{ij}\partial_j R \partial_i R=0\,, 
\label{ko11}
\eeq
so, we have proved the following \vspace{0.3cm}

\noindent {\bf Lemma}: In generic spherically symmetric space-times, the areal radius $R$ is conserved along Kodama trajectories.
\vspace{0.3cm}


In a generic spherically symmetric space-time a geometric dynamical identity holds true in general. This can be derived as follows. Let us introduce the normal space invariant 
\beq
\mathcal T^{(2)}=\gamma^{ij}T_{ij}\,.
\eeq 
which we shall term reduced stress energy tensor trace. Then, making use of Einstein equations, it is possible to show that, on the dynamical horizon, (see, for example, \cite{bob09})
\beq
\kappa_H=\frac{1}{2R_H}+2\pi R_H \mathcal T^{(2)}_H\,.
\label{vanzo}
\eeq
Introducing the horizon area and the (formal) three-volume enclosed by the horizon, with their respective differentials 
\beq
\mathcal A_H = 4\pi R_H^2\,,\qquad d \mathcal A_H=8\pi R_H dR_H\,,
\eeq
\beq
V_H=\frac{4}{3}\pi R_H^3\,,\qquad d V_H=4\pi R_H^2 dR_H\,,
\eeq
one gets
\beq
\frac{\kappa_{H}}{8 \pi}d \mathcal A_H =d\left(\frac{R_H}{2}\right) +\frac{ \mathcal T_H^{(2)}}{2} dV_H\,. 
\eeq
This equation can be re-cast in the form of a geometrical identity, once we introduce the Misner-Sharp energy at the horizon \cite{sean}: 
\beq
dM=\frac{\kappa_{H}}{2 \pi} d\left( \frac{\mathcal A_H}{4}\right)
-\frac{\mathcal T_H^{(2)}}{2} dV_H\,. 
\label{fl}
\eeq 

We conclude summarizing the main ingredients of the Hamilton-Jacobi tunneling method.
The Kodama vector, introduced above, gives a preferred flow of time and in this sense it
generalizes the flow of time given by the Killing vector in the static case.
As a consequence, we may introduce the invariant energy associated
with a particle of mass $m$ by means of the scalar quantity on the
normal space, the Kodama, or generalized Killing energy,
\beq 
\label{e} 
\omega =- \mathcal K^{i}\partial_i I\,, 
\eeq 
where $I$ is the particle action which we assume to satisfy the
reduced Hamilton--Jacobi equation 
\beq 
\label{hj} 
\gamma^{ij}\partial_i I \partial_j I + m^2=0\,. 
\eeq 
As we allow for non-minimal gravitational coupling, the substitution
$m^2 \rightarrow m^2+\xi \mathcal R $ is in order whenever $\xi\neq
0$, $\mathcal R$ being the Ricci curvature scalar and $\xi$ a
dimensionless coupling constant. 

If we are interested in Hawking effect, we may neglect the mass. One can reconstruct the action for particles coming
out of the horizon by
\begin{eqnarray}
I &=& \int dt\,\p_t I + \int dr\,\p_r I 
\end{eqnarray}
upon solving the Hamilton-Jacobi equation \eqref{hj} with zero mass. Then, one applies the near horizon approximation and
the null horizon expansion which forces us to make use of regular gauges across the horizon. 
These assumptions allow one to know the classical action and compute its imaginary part
making use of Feynman's prescription. The semiclassical result for the tunneling probability is then \cite{bob07,sean09,bob09}
\beq
\Gamma\simeq e^{-2 \Im I}\simeq e^{- \beta_H \omega_H}\,.
\eeq
In the static case, there is no doubt that $T_H=\frac{1}{\beta_H}$ can be interpreted as the Hawking temperature; but
what about the dynamical spherically symmetric case? From Einstein's equations and in presence of a dynamical horizon the geometrical dynamical Law holds as we have already seen. If also the Area Law for the black hole entropy is true, namely that 
\beq
S_H=\frac{\mathcal A_H}{4\hbar}\,,
\eeq
then, we can rewrite the geometrical identity (\ref{fl}) as a First Law of Thermodynamics for black holes, 
\beq
dM=T_H d S_H -\frac{T_H^{(2)}}{2} dV_H\,, 
\label{FL}
\eeq
with 
\beq
T_H=\hbar\,\frac{\kappa_H}{2\pi}\,. 
\eeq
Here we have re-introduced $\hbar$ in order to stress the quantum nature of entropy and temperature. Thus, it seems suggestive to interpret $T_H$ as a dynamical temperature of a slowly changing dynamical black hole, interpretation which is also supported by the fact that the Hayward dynamical surface gravity is an invariant for a generic spherically symmetric 
space-time. This interpretation has been put forward in \cite{sean09,bob09,bob07,peng}. Later in the paper, we shall try to present quantum theoretical arguments in order to substantiate this interpretation.

\section{Generic static black hole space-time}

As a first application of the formalism, let us consider a generic 
static black hole space-time. The starting point is a black hole metric in the Schwarzschild static gauge, 
\beq
ds^2=-V(r)dt^2+\frac{dr^2}{W(r)}+r^2 d \Omega^2 \;,
\label{s} 
\eeq
where, for sake of simplicity, we shall assume $W(r)=V(r)$, with $V(r)$ having just simple poles in order to describe what we might call a nice black hole. Let $r_H$ be the (greatest) solution of $V(r)=0$, the general formalism tells us that the horizon is located at $r=r_H$; the Kodama vector coincides with the usual Killing vector $(1,0,0,0)$; and the Hayward surface gravity is the Killing surface gravity, namely $\kappa_H=\kappa=\frac{V'_H}{2}$. 
This gauge is singular on the horizon, and it is not appropriate for the Hamilton-Jacobi tunneling method. For this reason we now introduce the Kruskal-like gauge associated with this static black hole solution. The first step consists in introducing the tortoise coordinate
\beq
dr^*=\frac{dr}{V(r)}\,.
\eeq
Then one has $-\infty < r^*< \infty$ and
\beq
ds^2=-V(r)dt^2+\frac{dr^2}{W(r)}+r^2 d \Omega^2=V(r^*)(-dt^2+(dr^{*})^2)+r^2(r^*) d \Omega^2 \;.
\label{s*} 
\eeq
Introduce Kruskal-like coordinates, according to
\beq
R=\frac{1}{\kappa}e^{\kappa r^*}\cosh(\kappa t)\,, \quad T=\frac{1}{\kappa}e^{\kappa r^*}\sinh(\kappa t)\,,
\eeq
one has
\beq
-T^2+R^2=\frac{1}{\kappa^2}e^{2 \kappa r^*}\,,
\label{ku}
\eeq
and
\beq
ds^2 &=& V(r^*)e^{-2\kappa r^*}(-dT^2+dR^2)+r^2(T,R) d \Omega^2 \nonumber \\
&\equiv& e^{\Psi(r*)}(-dT^2+dR^2)+r^2(T,R) d \Omega^2 \;,
\label{ks*} 
\eeq
where now the coordinates are $T$ and $R$, $r^*=r^*(T,R)$, $e^{\Psi(r*)}=V(r^*)e^{-2\kappa r^*}$, and the normal metric turns out to be conformally flat. In this gauge, the metric is a spherically symmetric
but time dependent one. The general formalism tells us that the horizon corresponds to 
\beq
(\partial_T\,r)_H=\pm (\partial_R \,r)_H\,,
\eeq
and this is equivalent to $T=\pm R$ and $r^* \rightarrow -\infty$. The Killing-Kodama vector is 
\beq
\mathcal K=e^{-\Psi(r^*)}\left(\partial_R \,r, -\partial_T\, r \right)\,. 
\eeq
Making use of the general formula, a direct, but tedious calculation confirms that the Hayward's surface gravity is still the Killing one. In fact, we have
\beq
\kappa_H=\frac{e^{-\Psi_H}}{2}\left( -\partial_T^2\, r+\partial_R^2\, r\right)_H=\frac{V'_H}{2}\,.
\eeq

Let us apply the Hamilton-Jacobi tunneling method, working in this time-dependent gauge. The Kodama energy is
\beq
\omega= e^{-\Psi(r^*)}\left(\partial_R \,r \partial_T\, I -\partial_T \,r \partial_R\, I \right)
\label{ke}
\eeq
where $I$ is the classical action. In this diagonal conformally flat gauge, the Hamilton-Jacobi equation is simply 
$\partial_T I=\pm \partial_R I$ and the null horizon expansion condition gives $dT=\pm dR$. 
Making the $+$ choice (outgoing particle), in the horizon approximation, one has
\beq
I\simeq 2 \int dR \partial_R I\,. 
\label{reconstr}
\eeq
Eq. (\ref{ke}) gives
\beq
\partial_R I=\frac{e^\Psi \omega}{\partial_R \,r-\partial_T \,r}\,. 
\eeq
The near horizon expansion gives
\beq
\partial_R \,r-\partial_T \,r \simeq (\partial^2_R r-\partial_T^2 r)_H(R-R_H)\,.
\eeq
so that the action (\ref{reconstr}) becomes,
\beq
I= \int dR \frac{\omega}{\kappa_H}\frac{1}{(R-R_H-i 0)}\,,
\eeq
Making use of Feynman's $i \epsilon$-- prescription for the simple pole, we finally get
\beq
\Im \, I=\frac{\pi \omega_H}{\kappa_H} =\frac{2 \pi\omega_H}{V'_H}\,.
\eeq
This imaginary part can be interpreted as arising because of a
non-vanishing tunneling probability rate of (massless) particles
across the event horizon, 
\beq
\Gamma \sim \exp\left(-2\Im \,I\right) \sim e^{-\frac{4 \pi}{ V'(r_H)} \, \omega_H}\, .
\eeq
The well-known result $T_H=\frac{V'_H}{4 \pi}$ is recovered.

\section{Cosmological horizons}

As a second application of the formalism, we consider a generic FRW
space-time with constant curvature spatial sections. Its line element
can be written as 
\beq
ds^2=-dt^2+a^2(t)\frac{dr^2}{1-\hat k r^2}+ [a(t)r]^2 d \Omega^2 \;.
\label{frwnf} 
\eeq
Here $\hat k :=\frac{k}{l^2}$, where $l$ is such that $a(t)l$ is
the curvature radius of 
the constant curvature spatial sections at time $t$ and, as usual,
$k=0, -1, +1$ 
labels flat, open and closed three--geometries, respectively. In this
gauge, the normal reduced metric is diagonal and 
\beq
\chi(t,r)=1- [a(t)r]^2\left[H^2(t) + \frac{\hat k}{a^2(t)}\right]\,.
\eeq
The dynamical trapping horizon is implicitly given by $\chi_H=0$, namely 
\beq 
R_H:= a(t) r_H = \frac{1}{\sqrt{ H^2(t)+\frac{\hat k}{a^2(t)}}}\,
,\qquad\mbox{with}\qquad H(t)=\frac{\dot a(t)}{a(t)}\;, 
\label{h}
\eeq
provided the space-time energy density $\rho(t)$ is positive. It
coincides with the Hubble radius as defined by astronomers for
vanishing curvature, but we shall call it Hubble radius in any
case. The dynamical surface gravity is given by equation (\ref{H})
and reads 
\beq
\kappa_H= - \left( H^2(t) +\frac{1}{2} \dot H(t) + \frac{\hat k}{2a^2(t)}\right)\,R_H(t) \,,\label{hfrw} 
\eeq
and the minus sign refers to the fact the Hubble horizon is, in
Hayward's terminology, of the inner type. According to (\ref{ko}), the
Kodama vector is 
\beq\label{kv}
\mathcal K= \sqrt{1-\hat k r^2}(\p_t -r H(t)\p_r )
\eeq 
so that the invariant Kodama energy of a particle is equal to 
\beq
\omega = \sqrt{1-\hat k r^2}(-\partial_t I + r H(t) \partial_r
I) \equiv \sqrt{1-\hat k r^2}\,\tilde{\omega}
\label{b2}
\eeq 
Notice that $\mathcal K$ is space-like for $ra>(H^2+\hat{k}/a^2)^{-1/2}$,
i.e. beyond the horizon. It follows that we can only ask for particles
to be emitted in the inner region $r< r_H$.\\
The next ingredient is the reduced Hamilton-Jacobi equation for a relativistic
particle with mass parameter $m$, 
\beq
-(\p_t I)^2 + \frac{(1-\hat k r^2) }{a^2(t)} \,(\p_r I)^2 + m^2=0\,.
\label{b}
\eeq
Making use of (\ref{b2}), one can solve for $\p_r I$, namely
\beq
\p_r I=-\frac{a H \tilde \omega (a r) \pm a\sqrt{\omega^2 -m^2 +
m^2\,\left(H^2 + \frac{\hat k}{a^2}\right)\,(a r)^2}}{1-\left(H^2
+ \frac{\hat k}{a^2}\right) \,(a r)^2}\,, 
\label{gg}
\eeq
with the signs chosen according to which direction we think the
particle is propagating.
The effective mass here defines two important and complementary energy
scales: if one is interested in the horizon tunneling then only
the pole matters (since the denominator vanishes), and we may neglect
to all the extents the mass parameter setting $m=0$ (since its
coefficient vanishes on the horizon). \\ 
On the opposite, in investigating other effects in the bulk away from
the horizon, such as the decay rate of composite particles, the 
role of the effective mass becomes relevant as the energy of the particle
can be smaller than the energy scale settled by $m$, and the square
root can possibly acquire a branch cut singularity. 

\subsection{The FRW space-time}

As an application of the last formula we may derive, following
\cite{bob09} (see also \cite{peng}), the cosmic 
horizon tunneling rate. To this aim, as we have anticipated, the
energy scale is such that near the 
horizon, we may neglect the particle's mass, and note that radially
moving massless particles follow a null direction. Then along a null
radial direction from the horizon to the inner region we have
\beq 
\Delta t = -\frac{a(t)}{\sqrt{1-\hat{k}r^2}} \Delta r. \label{ne}
\eeq
The outgoing particle action, that is the action for particles coming
out of the horizon towards the inner region, is then
\begin{eqnarray}
I &=& \int dt\,\p_t I + \int dr\,\p_r I \\
&=& 2 \int dr \p_r I
\end{eqnarray}
upon solving the Hamilton-Jacobi equation \eqref{b} with zero mass and using
(\ref{ne}). For $\p_rI$ we use now Eq.~(\ref{gg}), which exhibits a
pole at the vanishing of the function $F(r,t):=1-(a^2H^2+\hat k)r^2 $,
defining the horizon position. Expanding $F(r,t)$ again along a null
direction, one gets 
\beq
F(r,t) \approx + 4 \kappa_H a(t) (r-r_H) +\dots \;,\label{hay sg}
\eeq
where $\kappa_H$ is the Hayward dynamical surface gravity. 
In order to deal with the simple pole in the
integrand, we implement Feynman's $i\epsilon$~{--}~prescription. In
the final result, beside a real (irrelevant) contribution, we 
obtain the following imaginary part \cite{bob09} 
\beq \Im\, I =-\frac{\pi \omega_H}{\kappa_H}\, .\label{im} 
\eeq
This imaginary part is again interpreted as arising because of a
non-vanishing tunneling probability rate of (massless) particles
across the cosmological horizon, 
\beq
\Gamma \sim \exp\left(-2\Im \,I\right) \sim e^{-\frac{2 \pi}{(-
\kappa_H)}\, \omega_H}. 
\eeq
Notice that, since $\kappa_H <0$ and $\omega_H >0$ for physical
particles, (\ref{im}) is positive definite. As showed in \cite{bob09},
this result is invariant since the quantities appearing in the
imaginary part are manifestly invariant. Furthermore $T=-\kappa_H/2 \pi$ satisfies a First Law.
As a consequence, at least, in some asymptotic regime and for slowly changes in the geometry, we may
interpret $T=-\kappa_H/2 \pi$ as the dynamical temperature associated
with FRW space-times. 

In particular, this gives naturally a positive
temperature for de Sitter space-time, a long debated question years
ago, usually resolved by changing the sign of the horizon's energy.
It should be noted that in literature, the dynamical temperature is
usually given in the form $T=\frac{H}{2\pi}$ (exceptions are the
papers \cite{Wu:2008ir,chen}) with $H^{2}=\Lambda/3\equiv H_{0}^{2}$. Of course this is the expected result for de Sitter space in inflationary
coordinates, but it ceases to be correct in any other coordinate
system since, for example, $H=H_{0}\tanh H_{0}t$ in global coordinate system with positive spatial curvature. In this regard, the $\dot H$ and $\hat k$ terms are crucial in order to get an invariant temperature. Since this fact seems not so widely known, for sake of completeness, we shall try to show it in greater detail. 

de Sitter space in the global patch is described by the metric
\beq
ds^2=-dt^2+\cosh^2(H_0 t)\left[\frac{dr^2}{1-\hat k r^2}+ d\Omega^2\right] \;
\label{dsfrwnf} 
\eeq
with $a(t)=\cosh(H_0t)$, and $\hat k=H_0^2$. The Hubble parameter is time dependent
\beq
H(t)=H_0\tanh (H_0t)\,,
\eeq
and satisfies the identity
\beq
\dot H(t)=H_0^2-H^2(t)=\frac{\hat k}{a^2(t)}\,.
\eeq
Making use of (\ref{h}), the horizon radius is
\beq 
R_H:= a(t) r_H = \frac{1}{\sqrt{ H^2(t)+\frac{\hat k}{a^2(t)}}}=\frac{1}{H_0}\, 
\label{h0}
\eeq 
as it should be, and the Hayward's surface gravity
\beq
\kappa_H= - \left( H^2(t) +\frac{1}{2} \dot H(t) + \frac{\hat
k}{2a^2(t)}\right)\,R_H(t)=H_0 \,,
\label{hfrw0} 
\eeq
as it should, since it is an invariant quantity. Hence, we see that the $\dot H$ and $\hat k$ terms have to be present in a generic
FRW space-time. The important spatially flat case straightforwardly follows, 
\beq
\kappa_H= - \left( H(t) +\frac{\dot H(t)}{2 H(t)} \right) \,.
\label{hfrw1} 
\eeq
Note that this is independent on position, suggesting that $\kappa_{H}$ really is an intrinsic property of FRW space linked to the bulk. The horizon's temperature and the ensuing heating of matter was foreseen several years ago in the interesting paper \cite{Brout:1987tq}. We are going to challenge this interpretation.

\section{Quantum thermometers}

We recall that for the decay rate of a massive 
particle in de Sitter space, the exact quantum field theory calculation of Moschella et al.\cite{ugo} supports the WKB semiclassical 
tunneling result of \cite{Volo,bob10}. 

What about the other energy scale, the one associated with the horizon tunneling? We have shown that 
the semiclassical WKB method leads to an asymptotic particle production rate, involving a ``temperature''
\beq
\Gamma \simeq e^{-\frac{2 \pi}{|\kappa_H|} \omega_H}\quad \rightarrow T=\frac{|\kappa_H|}{2\pi}\,
\eeq
where $\kappa_H$ is the Hayward invariant surface gravity. For a generic spherically symmetric space-time, this result obtained by the Hamilton-Jacobi tunneling method seems a very clear prediction, namely an answer to the question: how hot is our expanding universe? 
A possible way to understand this issue using quantum field theory in curved space-time is to make use of a ``quantum thermometer'' (basically, an Unruh-DeWitt detector) and evaluate its response function, that is, loosely speaking, the number of clicks per unit proper time it detects as it is carried around the universe. For a recent review, see \cite{crispino}. 

In our approach, since we would like to obtain an invariant result, we will consider detectors which 
follow Kodama trajectories in a generic spherically symmetric space-time. The problem of back-reaction on these 
Kodama trajectories has been investigated in \cite{casadio}.

As we will see, the Unruh-DeWitt thermometer gives a clean answer only in the stationary case, and for FRW 
space-time the situation is not so simple, since horizon effects are entangled with highly non trivial kinematic effects. For general trajectories in flat space-time see the recent paper \cite{Kothawala:2009aj}.
 
An interesting analysis has been also put forward by Obadia \cite{obadia08}. 
In a recent paper \cite{nico}, local scaling limit techniques have been used 
in investigating the Hawking radiation.

\subsection{Unruh-DeWitt detector}

We recall that the Wightman function of a free, conformally coupled, massless scalar field (the only case we shall be interested in here) may be written as
\begin{equation}
W(x,x')=\sum_\alpha f_\alpha(x)f^*_\alpha(x')\;,
\label{w} 
\end{equation}
where the modes functions $f_\alpha(x)$ satisfy the equation 
\begin{equation}
\left(\Box -\frac{\mathcal R}{6}\right) f_\alpha(x)=0\;.
\label{f} 
\end{equation}
As it is well known, it is convenient to introduce the conformal time $\eta$ defined by
\begin{equation}
d \eta=\frac{dt}{a(t)}\;.
\label{ct} 
\end{equation}
Thus, the flat FRW space-time becomes conformally flat, 
\begin{equation}
ds^2=a^2(\eta)(-d\eta^2+d \vec x^2)\;.
\label{cf} 
\end{equation}
Let us denote by $x=(\eta, \vec x)$.
Making the ansatz 
\begin{equation}
f_{\vec k}(x)=\frac{g(\eta)}{a}e^{-i\vec k \cdot \vec x}\;,
\label{g} 
\end{equation}
one has, for the unknown quantity $g(k)$,
\begin{equation}
g''(\eta) + k^2 g(\eta) =0\;.
\label{g1} 
\end{equation}
As a consequence, making the choice of the vacuum given by
\beq
g(\eta)=\frac{e^{i\eta \vert k\vert }}{2\sqrt{\vert k\vert}}\,,
\eeq
one formally has 
\begin{equation}
W(x,x')=\int \frac{d \vec k}{2 k}\, \frac {e^{i(\eta-\eta')\vert k\vert -i\vec k \cdot (\vec x- \vec x')}}{a(\eta)a(\eta')}\;.
\label{w0} 
\end{equation}
This expression is meaningless as it stands and it has to be interpreted in the sense of distributions. 

Within Unruh-DeWitt detector issue, the usual prescription reads, \cite{BD},
\begin{equation}
W(x,x')=\lim_{\epsilon \rightarrow 0^+}\;\int \frac{d \vec k}{2 k}\, \frac {e^{i(\eta-\eta'-i\epsilon)\vert k\vert -i\vec k \cdot (\vec x- \vec x')}}{a(\eta)a(\eta')}\;.
\label{w2} 
\end{equation}
Integrating on $\vec k$, one arrives at 
\begin{equation}
W(x,x')= \lim_{\epsilon \rightarrow 0}\;\frac{1 }{4 \pi^2}\;\frac{1}{a(\eta)a(\eta')}\;\frac{1}{|\vec x- \vec x'|^2-|\eta-\eta'-i\epsilon|^2 }\;.
\label{wold} 
\end{equation}
However, it has been shown by Takagi \cite{T} and Schlicht \cite{S} that this prescription is non manifestly covariant. 
Since one is dealing with distributions, the limit $\epsilon \rightarrow 0^+$ has to be taken in the weak sense, otherwise
it may lead unphysical results with regard to instantaneous proper-time rate in Minkowski space-time for a Unruh-DeWitt detector.\\ Here, we adapt the Schlicht proposal (see also \cite{langlois,OM}) to our conformally flat case, namely
\begin{equation}
W(x,x')=\lim_{\epsilon \rightarrow 0^+} \int \frac{d \vec k}{2 k}\, \frac {e^{-i k ( x - x')-i\epsilon (\dot{x}+\dot{x}')}}{a(\eta)a(\eta')}\;,
\label{wS} 
\end{equation}
where an over dot is for derivative with respect to proper time. Integrating on $\vec k$, one arrives at 
\begin{equation}
W(x,x')=\lim_{\epsilon \rightarrow 0^+}\; \frac{1}{4 \pi^2}\;\frac{1}{a(\eta)a(\eta')}\;\frac{1}{[( x- x')+i\epsilon (\dot{x}+\dot{x}')]^2 }\;.
\label{w3} 
\end{equation}
In the flat case, this result has been generalized by Milgrom and Obadia, who made use of an analytical proper-time 
regularization \cite{OM}. It should be noted the appearance of Minkoswki contribution, as a function of the conformal time $\eta$. 

The transition probability per unit proper time of the detector depends on the response function per unit proper time
which for radial trajectories may be written as 
\beq 
\frac{d F}{ d\tau}=\lim_{\epsilon \rightarrow 0^+}\;
\frac{1}{2\pi^2}\,\mbox{Re}\int_{0}^{\tau -\tau_0} ds\, 
\frac{e^{-i E s}}{a(\tau)a(\tau-s)[x(\tau)-x(\tau-s)+i\epsilon(\dot{x}(\tau)+\dot{x}(\tau-s))]^2}\,,
\eeq
where $\tau_0$ is the initial time, and $E$ is the energy associated with the excited detector state. In some cases one can take the limit $\tau_0 \rightarrow -\infty$, but not in general since as a rule there is a Big Bang singularity. Instead we will be interested in the long term behavior of the detector, that is in the large $\tau$ limit at fixed $\tau_0$. The covariant $i\epsilon$-prescription is necessary in order to deal with the second order pole at $s=0$.
One may try to avoid the awkward limit $\epsilon \rightarrow 0^+$ by omitting the $\epsilon$-terms but subtracting the leading pole at $s=0$ (see for details \cite{louko}). In the calculation, the normalization condition
\beq
g_{\mu\nu}\dot{x}^{\mu}\dot{x}^{\nu}\equiv\left[a(\tau) \dot x(\tau)\right]^2= -1
\label{normaliz}
\eeq
characteristic of time-like four-velocities plays a crucial role. As a consequence, for $\Delta \tau >0$, after some 
calculations, one can present the detector transition probability per unit time in the form\footnote{This is not quite the original expression found in \cite{louko} but can be obtained from it by simple manipulations.}
\beq 
\frac{d F}{ d\tau}=\frac{1}{2 \pi^2}\int_{0}^{\infty}ds\, \cos( E s)
\left(\frac{1}{\sigma^2(\tau,s)} + \frac{1}{s^2}\right)-\frac{1}{2 \pi^2}\int_{\Delta \tau}^{\infty}ds\, 
\frac{\cos(E\,s)}{\sigma^2(\tau,s)}\,,
\label{L0}
\eeq 
where now there are no $i\epsilon$-terms and 
\beq
\sigma^2(\tau,s) \equiv a(\tau)a(\tau-s)[x(\tau)-x(\tau-s)]^2\,. 
\eeq
It should be noted the presence of the last fluctuating
finite time tail term. In some cases it controls how fast the thermal equilibrium is reached where, to recall the reader, $\Delta\tau$ is the duration of the experiment, i.e. the time the detector is switched on. 

We can go further, observing that due to (\ref{normaliz}), for small $s$, one has
\beq
\sigma^2(\tau,s)=-s^2[1+O(s^2)]\,.
\eeq
Thus we may write
\beq
\sigma^2(\tau,s)=-s^2[1+s^2d(\tau,s)]\,.
\eeq
As a result, we may introduce the even $\sigma_e^2(\tau,s)$ and odd part $\sigma_o^2(\tau,s)$ in $s$ of the 
quantity $\sigma^2(\tau,s)$ and arrive at the two contributions
\beq 
\frac{d F}{ d\tau}=\frac{d F_e}{ d\tau}+\frac{d F_o}{ d\tau}-\frac{1}{2 \pi^2}\int_{\Delta \tau}^{\infty}ds\, 
\frac{\cos (E\,s)}{\sigma^2(\tau,s)} \,,
\label{L}
\eeq
where
\beq 
\frac{d F_e}{ d\tau}= \frac{1}{4 \pi^2}\int_{-\infty}^{\infty}ds\, \cos( E s)
\left(\frac{1}{\sigma_e^2(\tau,s)} + \frac{1}{s^2}\right)\,,
\label{Le}
\eeq 
\beq 
\frac{d F_o}{ d\tau}=\frac{1}{2 \pi^2}\,\int_{0}^{\infty}ds\, \frac{\cos( E\,s)}{\sigma_o^2(\tau,s)}\,.
\label{Lo}
\eeq
This is the main formula which we will use in the following. One can view the pole subtraction as an elementary example of a renormalization procedure, one that would not be necessary where the $\epsilon$-terms were kept all along. However, Eq.s~\eqref{L0}, \eqref{L} are much more convenient to deal with than the original expression containing the $\epsilon$-terms, since the limit in distributional sense must also be taken at the end of any computation.\\
In the important stationary case in which $\sigma(\tau,s)^2=\sigma^2(s)=\sigma^2(-s)$, for $E >0$ and $\Delta \tau >0 $ , 
the odd part drops out and one simply has
\beq 
\frac{d F}{ d\tau}=\frac{1}{4\pi^2}\,\int_{-\infty}^{\infty}ds\, e^{-i E s}\left( \frac{1}{\sigma^2(s)}+\frac{1}{s^2}\right)
-\frac{1}{2 \pi^2}\int_{\Delta \tau}^{\infty}ds\, 
\frac{\cos(E\,s)}{\sigma^2(s)} \,. \label{Lr}
\eeq 
In these cases, examples are the FRW de Sitter space and static black holes, we shall take the limit $\Delta\tau\to\infty$, and the fluctuating tail vanishes.

\subsection{The generic static black hole revisited}

It is clear that the Unruh-DeWitt detector formalism can be applied to the static black hole of Section 3. The key point to recall here is that in the Kruskal gauge (\ref{ks*}) the normal metric is conformally related to two dimensional Minkoswki space-time, and the normal metric is the important one for radial trajectories. The second observation is that the the Kodama observers are defined by the integral curves associated with the Kodama vector, thus the areal radius $r(T,R)$ is {\bf constant}, and $r^*$ is constant. As a consequence, the proper time along Kodama trajectories reads
\beq
d\tau^2=V(r^*)dt^2=e^{\Psi(r^*)}(dT^2-dR^2)=a^2(r^*)(dT^2-dR^2)\,,
\eeq
so that
\beq
t=\frac{\tau}{\sqrt{V(r^*)}}\,, 
\eeq
and 
\beq
R(\tau)=\frac{1}{\kappa}e^{\kappa r^*}\cosh\left(\kappa\frac{\tau}{\sqrt{V(r^*)}}\right) \,,\qquad T(\tau)=\frac{1}{\kappa}e^{\kappa r^*}\sinh \left(\kappa \frac{\tau}{\sqrt{V(r^*)}} \right)\,.\label{cazzo}
\eeq 
Recalling that the geodesic distance is
\beq
\sigma(\tau,s)=e^{\Psi(r^*)}\left[ -\left(T(\tau)-T(\tau-s)\right)^2+\left(R(\tau)-R(\tau-s)\right)^2 \right]\,,
\eeq
one gets for $T(\tau)$ and $R(\tau)$ of the form (\ref{cazzo}) 
\beq\label{figa}
\sigma^2(\tau,s)=-\frac{4 V(r^*)}{\kappa^2} \sinh^2\left(\frac{\kappa\,s}{2\sqrt{V(r^*)}}\right)\,. \label{sinh2}
\eeq
Since $\sigma^2(\tau,s)= \sigma^2(s) = \sigma^2(-s)$, we can use the formula (\ref{Lr}) in the limit $\tau $ goes 
to infinity:
\beq 
\frac{d F}{ d\tau}=
\frac{\kappa}{8\pi^2 \sqrt{V^*}}\int_{-\infty}^{\infty}dx e^{-i\frac{2 \sqrt{V^*} E x}{\kappa}}\left( -\frac{1}{\sinh^2 x}+\frac{1}{x^2}\right)\,,
\eeq 
The integral can be evaluated by the Theorem of Residues and the final result is
\beq 
\frac{d F}{ d\tau}= \frac{1}{2\pi}\; \frac{E}{e^{\frac{2\pi \sqrt{V^*} E}{\kappa}}-1}\,.
\eeq
Since the transition rate exhibits the characteristic Planck distribution, it means that the Unruh-DeWitt thermometer 
in the generic spherically symmetric black hole space-time 
detects a quantum system in thermal equilibrium at the local temperature 
\beq
T=\frac{\kappa}{2\pi \sqrt{V^*}}\,.
\label{z}
\eeq
With regard to the the factor $\sqrt{V^*}=\sqrt{-g_{00}}$, recall the Tolman's Theorem, which states that for gravitational system at 
thermal equilibrium, one has {$T\sqrt{-g_{00}}=\mbox{Constant} $. For asymptotically flat black hole space-times, one obtains the ``intrinsic'' constant temperature of the Hawking effect, i.e.
\beq
T_H=\frac{\kappa}{2\pi}= \frac{V'_H}{4\pi}\,.
\eeq

It is not necessary to stress how this result agrees with several derivations of Hawking effect, including the tunneling approach we have already discussed. 

We would like to point out that this is a quite general result, valid for a large class of nice black holes, as for example
Reissner-Nordstr\"{o}m and Schwarzschild-AdS black holes. On the other hand, the Schwarzschild-dS black hole cannot be 
included, due to the presence of two horizons. However, as an important particular case, we may consider the 
static de Sitter black hole, defined by
\beq
V(r)=1-H_0^2r^2\,, \qquad H_0^2=\frac{\Lambda}{3}\,.
\eeq
The unique horizon is located at $r_H=\frac{1}{H_0}$ and Gibbons-Hawking temperature is \cite{GH} 
\beq
T_H=\frac{H_0}{2\pi}\,.
\eeq
We will present another derivation of this well known result in another gauge in the next Section.

We conclude this Section, making some remarks on de Sitter and Anti-de Sitter black holes. First, we observe that in a static space-time, namely the one corresponding to a nice black hole, the Killing-Kodama observers with $r=K$ constant, have an invariant acceleration 
\beq
\mathcal A^2=g_{\mu\nu}\mathcal A^\mu \mathcal A^\nu=\frac{V'^2(K)}{4 V(K)}\,,
\label{a}
\eeq
where $\mathcal A^\mu = u^\nu \nabla_{\nu} u^\mu$, $u^\mu$ being the observers 4-velocity. In the case of de Sitter black hole, one has
\beq
\mathcal A^2=\frac{H_0^4 K^2}{1-H_0^2 K^2}\,.
\eeq
As a result, 
\beq
\mathcal A^2+H_0^2=\frac{H_0^2}{1-H_0^2 K^2}\,,
\eeq
and the de Sitter local temperature felt by the Unruh detector, 
\beq
T_{dS}=\frac{H_0}{2\pi}\frac{1}{\sqrt{1-H_0^2 K^2}}
\eeq
can be re-written as \cite{thirr,deser98}
\beq
T_{dS}=\frac{1}{2\pi}\sqrt{\mathcal A^2+H_0^2}=\sqrt{T^2_U+T^2_{GH}}\,.
\eeq
A similar result was also obtained for AdS in \cite{deser98}, and it reads
\beq
T_{AdS}=\frac{1}{2\pi}\sqrt{\mathcal A^2-H_0^2}\,.
\label{d}
\eeq
We would like to show that it is again a particular case of our general formula (\ref{z}). In fact, it is sufficient to 
apply it to the four-dimensional topological black hole with hyperbolic horizon manifold found in 
\cite{Beng,Mann,Brill,Vanzo}, which is a nice black hole with
\beq
V(r)=-1-\frac{C}{r}+H_0^2r^2\,,
\eeq
where $C$ is a constant of integration related to the mass. The space-time is a solution of Einstein equation with
negative cosmological constant $\Lambda=-3H^2_0$, which is asymptotically Anti-de Sitter. When the constant of 
integration goes to zero, one has still a black hole solution, and calculation similar to the one valid for de Sitter space-time gives
\beq
T_{AdS}=\frac{H_0}{2\pi}\frac{1}{\sqrt{-1+H_0^2 K^2}}=\frac{1}{2\pi}\sqrt{\mathcal A^2-H_0^2}\,,
\eeq 
which is the Deser \textit{et al.} result. Thus, for symmetric space time with constant curvature one has that the local temperature
felt by the Kodama-Unruh-DeWitt detector can be written as 
\beq\label{UH}
T=\sqrt{T^2_U + \varsigma T^2_{GH}}\,,
\label{z2}
\eeq 
where $T_U$ is the Unruh temperature associated with the acceleration of the Kodama observer, $T_{GH}$ is the 
Gibbons-Hawking temperature and $\varsigma=1$ for the de Sitter space-time, $\varsigma =0$, for Minkowski space-time (this is the original Unruh
effect) and $\varsigma=-1$ for the ``massless" AdS topological black hole. This formula may help to understand better the relation between the Unruh-like effects and the genuine presence of a thermal bath and shows that, in general, the Kodama-Unruh detector gives an intricate relation between Killing-Hayward temperature and other invariant temperatures such as the Unruh's one. 

\section{The FRW and de Sitter expanding universes}

To apply the Unruh-DeWitt detector formalism to cosmology we consider a generic FRW spatially flat space-time.
This case has been investigated also in \cite{obadia08}. Recall that here the areal radius is $R=ra(t)$. Thus, for the Kodama observer, one has
\beq
r(t)=\frac{K}{a(t)},
\eeq
with constant $K$. For a radial trajectory, the proper time in FRW is
\beq
d\tau^2=a(\eta)(d\eta^2-dr^2)\,.
\eeq
As a function of the proper time, the conformal time along a Kodama trajectory is
\beq
\eta(\tau) &=& -\int d \tau \,\frac{1}{a(\eta) \sqrt{1-K^2H^2(\tau)}} d \eta \nonumber \\
&\equiv& -\int d\tau\,\frac{1}{a(\tau)\sqrt{V(\tau)}} \;,
\eeq
$H(\tau)$ being the Hubble parameter as a function of proper time. In general, we may use Equation (\ref{L}) in which, for radial Kodama observer, one has
\beq
x(\tau)=(\eta(\tau), r(\tau),0,0)=\left(-\int \frac{1}{a(\tau)\sqrt{V(\tau)}} d \tau, \frac{K}{a(\tau)},0,0\right)\,.
\eeq 

A very important example of FRW space is the stationary flat de Sitter expanding (contracting) space-time, which, in the FRW contest, is 
defined by $a(t)=e^{H_0t}$. Thus, 
\beq
ds^2=-dt^2+e^{2H_0t} (dr^2+r^2d\Omega^2)\,.
\eeq
Here $H(t)=H_0$ is constant as well as $V=V_0=1-H_0^2K^2$. For Kodama observers 
\beq
\tau=\sqrt{V_0}\, t\,, \quad \quad a(\tau)=e^{\frac{H_0}{\sqrt{V_0}}\tau}\,,
\eeq
and
\beq
\eta(\tau)= -\frac{1}{H_0}e^{-\frac{H_0}{\sqrt{V_0}}\tau}\,, \quad r(\tau)= K \,e^{-\frac{H_0}{\sqrt{V_0}}\tau}\,, 
\eeq
so, the geodesic distance is
\beq\label{tetta}
\sigma^2(\tau,s)=-\frac{4\,V_0}{H_0^2}\sinh^2\left(\frac{H_0 \,s}{2\sqrt{V_0}}\right)\,.
\eeq
This result is formally equal to the one obtained for the generic static black hole (\ref{sinh2}). 
Since again $\sigma^2(\tau, s)= \sigma^2(s)=\sigma^2(-s)$, we may use (\ref{Lr}) and obtain, for $E >0$ and in the infinite time limit
\beq\label{desitt}
\frac{d F}{ d\tau}=\frac{H_0}{8\sqrt{V_0}\pi^2}\,\int_{-\infty}^{\infty}d x\, e^{-i\frac{2 \sqrt{V_0} E x}{H_0}}\left( -\frac{1}{\sinh^2 x}+\frac{1}{x^2}\right)\,.
\eeq 
Again, we arrive at
\beq 
\frac{d F}{ d\tau}=\frac{1}{2\pi} \,\frac{E}{e^{\frac{2\pi \sqrt{V_0} E}{H_0}}-1}\,,
\eeq
which shows that the Unruh-DeWitt thermometer in the FRW de Sitter space detects a quantum system in thermal equilibrium at a temperature $T=\frac{H_0}{2\pi \sqrt{V_0} }$. Here, the Tolman factor is substituted by a Lorentz factor, which represents, as we already know, the Unruh acceleration part. In fact we recall that 4-acceleration of a Kodama observer in a FRW space-time turns out to be
\beq
\mathcal A^2 = \mathcal A^\mu \mathcal A_\mu = 
K^{2}\left[\frac{\dot H(t) + (1- H^2(t) K^2) H^2(t)}{(1 - H^2(t) K^2)^\frac{3}{2}}\right]^2
\eeq
where $\mathcal A^\mu := u^\nu \nabla_\nu u^\mu$, $u^\mu$ being the 4-velocity of the detector, that is the (normalized) tangent vector to the integral curves of the Kodama vector field. As a result, for dS space in a time dependent patch we have 
\beq
\mathcal A^2=\frac{K^2H_0^4}{1-K^2H_0^2}\,,
\eeq
showing that 
\[
\frac{H_{0}}{\sqrt{1-H_{0}^{2}K^{2}}}=\sqrt{H_{0}^{2}+\mathcal{A}^{2}}
\]
in agreement with the dS static calculation. When $K=0$, one has $V_0=1$ and the classical 
Gibbons-Hawking result $T_{dS}=\frac{H_0}{2\pi}$ is recovered. 

We close this section with a brief discussion of an asymptotic phenomenon which will be relevant to the following sub-section, first noted in de Sitter space by Garbrecht, Prokopec \cite{proco}: how is the thermal distribution of the response function reached in the limit of very large times? To this aim we consider the finite time contribution due to the fluctuating tail (the last term in Eq.~\eqref{Lr}) for 
de Sitter or Black hole cases. A direct calculation of the tail using Eq.~\eqref{figa} for black holes or Eq.~\eqref{tetta} for dS, gives

\beq\label{tailds}
\frac{\kappa^2}{8\pi^2}\,\int_{\Delta \tau}^{\infty}d s\, \frac{\cos (E\,s)}{\sinh^2 \frac{\kappa s}{2}}=
\frac{\kappa}{2\pi^{2}}\sum_{n=1}^{\infty}\frac{n e^{-2\pi nT_{H}\Delta \tau}}{n^2+\frac{E^2}{4\pi^{2}T_{H}^{2}}}\left(\frac{E}{2\pi T_{H}}
n\cos(E \Delta \tau)-\sin (E \Delta\tau)\right)
\eeq
where $\kappa$ is the horizon surface gravity and $T_{H}$ is the local Hawking temperature of the horizon, given by $\kappa/2\pi\sqrt{V}$. We recall that $\kappa=H_{0}$ for de Sitter space and $V^{'}(r_{H})/2$ for the black hole.\\ 
As a result, the fluctuating tail term drops out exponentially. For large $E/T_H$ the oscillations are at first very large relative to the equilibrium value and for few Hubble times then they decay moderately fast, reaching values comparable to equilibrium values only at times $\Delta\tau\sim\;\,\mbox{a\;few}\; E/T_{H}$. For instance with $E/T_{H}=15$ we have a plot\footnote{Obtained using Wolfram's Mathematica numerical code.}

\begin{figure}[ht]
\includegraphics{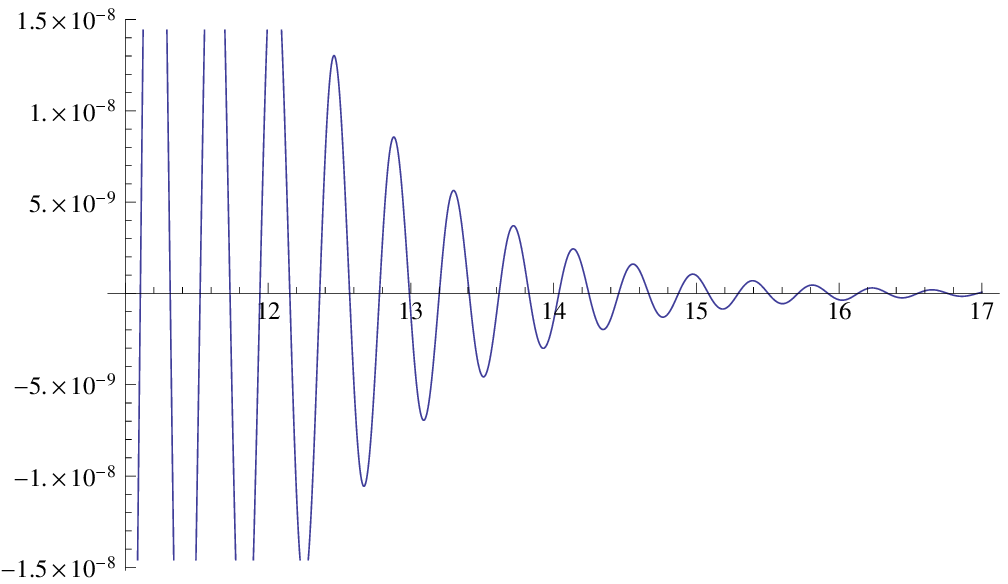}
\end{figure} 
although only at $\Delta\tau\simeq90$ is the tail smaller than the equilibrium value, which if of the order $10^{-41}$. From this point of view the oscillating terms seem more sensible than the equilibrium value.

\subsection{Asymptotically de Sitter space-times}

We end with some considerations about the non stationary case which appear when space-time is only spatially homogeneous as in generic FRWL models, but limiting our computations to comoving detectors (i.e. $K=0$), the less interesting from the point of view of the Unruh effect. 
It should be clear from previous considerations, that the formula for the transition rate of the detector (\ref{L0}) is manageable, in the sense that we can extract quantitative information, only in few highly symmetrical circumstances. For instance the case of a realistic cosmology dominated by matter and vacuum energy has a scale factor of the form
\beq
a(t)=a_{0}\sinh\left(\frac{3}{2}\sqrt{\Omega_{\Lambda}}H_{0}t\right)^{2/3}, \qquad a_{0}=\left(\frac{\Omega_{m}}{\Omega_{\Lambda}}\right)^{1/3}
\eeq
where $H_{0}$ is the present value of the Hubble constant. We put $h=3\sqrt{\Omega_{\Lambda}}H_{0}/2$ for simplicity: then the conformal time is given up to a constant by the hypergeometric function
\beq
\eta(t)=\int\frac{dt}{a(t)}=-\frac{1}{a_0 h}\,\sech^{2/3}(ht)\,_{2}F_{1}(5/6,1/3;4/3;\sech^{2}(h t)) +C
\eeq
where $C$ is possibly a complex constant but the first term is real. Clearly integrals such as \eqref{L0}, \eqref{Le} and \eqref{Lo} are now outside an analytical treatment and one can resort to numerical evaluations. We only note few things: at small $ht\ll1$ the model follows the Einstein-de Sitter law $a(t)\sim t^{2/3}$ of a matter dominated universe. In this case the geodesic distance is
\[
\sigma^{2}(\tau,s)=6\tau(\tau-s)\sinh^{2}\left[\frac{1}{6}\log\left(1-\frac{s}{\tau}\right)\right]
\]
and the pole structure is the following: $s=\tau$ with residue $0$ (also a branch point) and $s=0$, a double pole with residue $-2iE$. So this detector will not take notice of the expansion of the universe, even if the horizon temperature is non vanishing ($T_{H}\sim t^{-1}$). Second, there will be in general further poles over those along the imaginary axis, due to the factor $a(\tau-s)$ which multiplies the geodesic distance and which may have zeroes when extended throughout the complex plane: for example with $a(\tau)$ as given we have further poles at
\[
s_{n}=\tau+\frac{n\pi i}{h}
\]
which however seem to disappear in a large time limit. Moreover, we have some numerical evidence that their effect is eaten by the oscillating tail that we discussed in the context of dS space. Finally, we have the noteworthy feature that the geodesic distance appearing as integrand in the response function is time dependent, which means the detector will not feel a thermal bath except asymptotically at large Hubble times. In fact at large $ht\gg1$ the scale factor describes de Sitter space with the smaller Hubble constant $\sqrt{\Omega_{\Lambda}}H_{0}$ and $\sigma^{2}(\tau,s)$ exhibits the characteristic dS double poles at $s_{n}=2n\pi i/\sqrt{\Omega_{\Lambda}H_{0}}$, so we expect an asymptotic  thermal distribution with $T= \sqrt{\Omega_{\Lambda}H_{0}}/2\pi$. \\
This can be seen in few simple toy models, which are not physically motivated, but that allow some analytical evaluation. For instance one could try the scale factor $a(t)$ (always with spatial curvature $k=0$)
\begin{equation}
a(t) = \cosh^2\left(\frac{H_0}{2} t\right)\,. 
\end{equation}
This scale factor tends to a de Sitter expansion phase at late times and it is singularity free. The Unruh--DeWitt detector is understood to sit at the  origin of the comoving mesh so we have the huge simplification
\beq
\tau(t) = t.
\eeq
As a consequence, the conformal time is 
\beq
\eta(t) =-\frac{2}{H_0}\tanh \left(\frac{H_0}{2}t\right)\,,
\eeq
and the geodesic distance reads
\beq
\sigma^2(\tau,s)= - a(\T) a(\T-s) \left[\eta(\T) - \eta(\T-s) \right]^2=
-\frac{4}{H_0^2}\sinh^2 \left(\frac{H_0}{2}t\right)\equiv \sigma_{dS}^2(s)\,.
\eeq
Clearly the transition probability rate is given by the equation (\ref{L}), for $E>0$, and it is formally equal to the de Sitter case. We stress that this is only true for Kodama trajectories, $r(t)=K/a(t)$, with $K=0$. For non vanishing $K$, $\eta(\tau)$ may be expressed as an intractable elliptic integral, and the transition probability rate is time dependent. The same result can be obtained by the choice 
\begin{equation}
a(t) = \sinh^2\left(\frac{H_0}{2} t\right)\,, 
\end{equation}
this toy model being asymptotically de Sitter, but close to Milne model and singular near $t=0$. 

The following toy models are less trivial but, again for $K=0$, they exhibit features which are generally present for more realistic models. They are described by the choices
\begin{equation}
a(t) = \frac{1}{ \cosh (H_0 t)}\,, 
\end{equation}
and
\begin{equation}
a(t) = \frac{1}{ \sinh (H_0 t)}\,. 
\end{equation}
These models represent asymptotically contracting dS spaces, and their geodesic distance reads respectively
\beq
\sigma^2(\tau,s)= \sigma^2_{dS}(s)\frac{\cosh^2 H_0(t-s/2)}{\cosh H_0t \cosh H_0(t-s)}\,,
\eeq
\beq
\sigma^2(\tau,s)= \sigma^2_{dS}(s)\frac{\sinh^2 H_0(t-s/2)}{\sinh H_0t \sinh H_0(t-s)}\,.
\eeq
We will investigate only the first one. Using trigonometric identities one can write
\beq
\frac{1}{\sigma^2(\tau,s)}=\frac{1}{\sigma_{dS}^2(s)}-\frac{H_{0}^{2}}{4\cosh^{2}(H_{0}(t-s/2))}
\eeq
Thus the response function is the de Sitter one, which we already discussed, plus the integral
\beq
 J=-\frac{H_{0}^{2}}{8\pi^{2}}\int_0^{\Delta t}\frac{\cos Es}{\cosh^{2}(H_{0}(t-s/2)}\,ds
 \eeq
Using
\[
\sech^{2}(x)=-4\sum_{n=1}^{\infty}(-1)^{n}n\,e^{-2nx}
\]
a term by term integration gives
\beq\label{tailzer}
J=\frac{H_{0}}{2\pi^{2}}\sum_{n\geq1}(-1)^{n}\frac{n\,e^{-nH_{0}\Delta t}}{n^{2}+\frac{E^{2}}{H_{0}^{2}}}\,\left[n(\cos E\Delta t-1)+\frac{E}{H_{0}}\,\sin E\Delta t\right]
\eeq
The similarity with de Sitter case, Eq.~\eqref{tailds}, is quite evident so some of the features of the pure dS case are indeed present. 
To summarize, we would say that the detector clicks close to a de Sitter response and reaches thermalization through decaying oscillations as  $\tau$ goes to infinity. In fact, as far as the regime $H_0 \T \gg 1$ is concerned, expression (\ref{tailzer}) simplifies to the de Sitter space result. We may think of this as describing a de Sitter thermal noise continuously perturbed by the expansion (or contraction) of the universe. In particular, insofar as we can speak of temperature, it registers the de Sitter temperature, different from the horizon temperature parameter which in the present case is $2\pi T_{H}=H_{0}-2H_{0}\sech(2H_{0}t)$. This illustrates the feature that comoving detectors are probably unable to reveal the horizon temperature, whose quantum interpretation as an Unruh effect requires non trivial Kodama trajectories. But for such accelerated detectors ($K \neq 0$) one expects even more complicated results as Eq.~\eqref{UH} may lead one to think.

\section*{Conclusions}

In this paper, making use of the Hamilton-Jacobi tunneling method, we have re-derived the Hawking effect for a generic static black hole and the analogous tunneling phenomenon which appear to exist also in a generic FRW space-time possessing a trapping horizon. The method also works for slowly changing black holes, except that the event horizon gets replaced by a time-like surface known as the Hayward's trapping horizon. With the aim to better understand the temperature-versus-surface gravity paradigm, the asymptotic results obtained by this semiclassical method have been tested with more reliable quantum field theory techniques as the Unruh-DeWitt detector analysis. For black holes the two analysis are mutually consistent and even predict the dependence of the temperature on position or acceleration. For cosmology and away from de Sitter space the thermal interpretation is lost but the detector still gets excited by the expansion of the universe: the second set of poles, as well as the presence of an odd part, arising in a quasi-de Sitter universe signals this lack of thermal behavior. Note, this result seems to show how the thermal interpretation breaks down in most of the cases: the time-dependence of the transition rate is expected to be persistent in all purely non-de Sitter dynamical solutions. Moreover, from the comoving detector point of view the horizon surface gravity seems less prominent and not unambiguously associated with a temperature parameter as the tunneling method suggests. For instance, in the Einstein-de Sitter regime there seems to be no excitations (for co-moving detectors only). It remains to see whether there is any non trivial effect on accelerated Kodama, or more general, trajectories. \\
The fact that our co-moving monopole detector apparently seems unaware of the trapping horizon is related to the ambiguity of the particle concept in cosmology and forces, in our view, a different interpretation of the tunneling picture, hopefully giving matter to future work. One possibility is that the horizon surface gravity could represent an intrinsic property of the horizon itself, leading to some kind of holographic description, while the detector in the bulk simply clicks because it is immersed in a changing geometry. In fact, we would expect the clicks in almost any changing geometry, even for those lacking a trapping horizon.

\ack{We thank S.~Hayward and G.~Cognola for useful discussions.}

\end{document}